# Deposition Order Dependent Magnetization Reversal in Pressure Graded Co/Pd films

P. K. Greene,[1] B. J. Kirby,[2] J. W. Lau,[2] J. A. Borchers,[2] M. R. Fitzsimmons,[3] and Kai Liu[1,*]

[1]*Physics Department, University of California, Davis, CA 95616*

[2]*National Institute of Standards and Technology, Gaithersburg, MD 20899*

[3]*Los Alamos National Laboratory, Los Alamos, NM 87545*

Abstract

Magnetization reversal mechanisms and depth-dependent magnetic profile have been investigated in Co/Pd thin films magnetron-sputtered under continuously varying pressure with opposite deposition orders. For samples grown under increasing pressure, magnetization reversal is dominated by domain nucleation, propagation and annihilation; an anisotropy gradient is effectively established, along with a pronounced depth-dependent magnetization profile. However, in films grown under decreasing pressure, disorders propagate vertically from the bottom high-pressure region into the top low-pressure region, impeding domain wall motion and forcing magnetization reversal via rotation; depth-dependent magnetization varies in an inverted order, but the spread is much suppressed.

Recent advances in nanomagnetics have fundamentally changed information technology and our everyday life. Today's compact hard disk drives (HDD's) are already approaching an areal density of 1 Terabits/in$^2$ and cost only ~6¢/GigaByte.[1-3] However, continued advancement of magnetic recording is facing multiple correlated challenges in signal-to-noise ratio, thermal stability, and writeability.[4] There has been an intense pursuit of ideal ultrahigh density media that can meet these challenges, particularly in reconciling competing requirements for thermal stability and writeability. Several approaches have been proposed, including energy-assisted magnetic recording (heat[1-3, 5, 6] or microwave[7, 8]), bit-patterned media,[9-11] or magneto-optical recording.[12, 13] An alternative, technologically less disruptive, approach is to use exchange composite[14, 15] or graded magnetic anisotropy media[16, 17] where the magnetic anisotropy varies along the depth of the film. In such media the interlayer exchange coupling lowers the overall coercivity, facilitating the writing process, while the magnetically hard layer provides pinning for the media and ensures its thermal stability. Typically the anisotropy gradient has been introduced by varying synthesis conditions[18, 19] or individual layer thicknesses,[20] introducing composition gradients,[21, 22] or inducing structural variations.[23, 24]

An important effect that has so far been largely unexplored is the deposition order dependence of graded anisotropy, i.e., having the high anisotropy component on top of the low anisotropy one, or *vice versa*. The seemingly equivalent structures are inversion asymmetric, due to the substrate, and thus would impact the actual magnetic recording process. Effects of disorder propagation, strain relaxation, or potentially the Dzyaloshinskii–Moriya interaction[25, 26] could all influence the magnetic characteristics, particularly along the depth of the media which are difficult to resolve. To cleanly investigate such deposition-order dependence requires model systems with opposite anisotropy gradients and a comprehensive set of structure and magnetic probes that are sensitive to local as well as extended length scales, both along the film depth and in the film plane.

In this work we have accomplished this task and investigated the effect of deposition order in graded anisotropy Co/Pd films. By varying the magnetron sputtering pressure during deposition we have induced structural and magnetic anisotropy gradients in a controlled fashion. Using x-ray diffraction (XRD), transmission electron microscopy (TEM), polarized neutron reflectometry (PNR), and the first-order reversal curve (FORC) method we have probed the depth-dependent structure and magnetization profiles as a function of both the deposition pressure and the deposition order of the layers. We report the influence of sputtering pressure on local layer structure, the propagation of disorder vertically within the films, and the effects on magnetization reversal mechanisms and depth-dependent magnetic profile.

Thin film samples were grown by magnetron sputtering onto Si (100) substrates in a vacuum chamber with a base pressure of $9\times10^{-9}$ torr. A 20 nm Pd seed layer was first deposited at an Ar sputtering pressure of 0.7 Pa. Then, $[Co(0.4nm)/Pd(0.6nm)]_{60}$ multilayers were grown with the following conditions and capped with 4.4 nm of Pd sputtered at 0.7 Pa:

1) Low Pressure (LP): all 60 Co/Pd bilayers sputtered at 0.7 Pa.

2) 3 Pressure 1 (3P1): bottom 30 bilayers sputtered at 0.7 Pa, the next 15 bilayers at 1.6 Pa and the top 15 bilayers at 2.7 Pa.

3) 3 Pressure 2 (3P2): bottom 15 bilayers sputtered at 2.7 Pa, the next 15 bilayers at 1.6 Pa, and the top 30 bilayers at 0.7 Pa.

4) High Pressure (HP): All 60 Co/Pd bilayers sputtered at 2.7 Pa.

As shown in prior studies, varying the sputtering pressure during deposition is a convenient way to tailor the interface and magnetic anisotropy in perpendicular anisotropy systems.[18, 19, 27] At low Ar pressures the sputtered atoms arrive at the substrate with sufficient kinetic energy and mobility, leading to relatively pristine layer structures with smooth interfaces;[28] at higher pressures, the sputtered atoms have much less kinetic energy, resulting in rougher interfaces[29] and more disorders (e.g., distinct grain boundary phase formation).[30]

Structural characterization was performed by low angle x-ray reflectometry (XRR) and high angle XRD using a Bruker D8 thin film x-ray diffractometer with Cu K$_\alpha$ radiation. TEM was performed in both cross-sectional and plan view geometries to investigate the film microstructure. Magnetic properties were studied by magnetometry and the FORC method[31-33] using a Princeton Measurements Corp. vibrating sample magnetometer (VSM). For FORC studies, the sample is brought from positive saturation to a reversal field ($H_r$), and the magnetization $M$ is measured as the applied field $H$ is increased to positive saturation. The FORC distribution is calculated as a second order mixed partial derivative of $M$ ($H, H_r$): $\rho \equiv -\partial^2 M(H, H_r)/2\partial H \partial H_r$, which is sensitive to irreversible switching events,[34, 35] providing "fingerprints" of different magnetization reversal processes.[36] The depth-dependent magnetic profile was studied by PNR using the Asterix reflectometer / diffractometer at the Los Alamos Neutron Science Center and the NG1 Reflectometer at the NIST Center for Neutron Research. PNR is sensitive to the depth profiles of the nuclear composition and in-plane magnetization of thin films and multilayers, as demonstrated in prior studies of graded media.[18, 20, 21]

XRR measurements show clear oscillations for all 4 samples, indicating overall good layer structures (Fig. 1a). The oscillation amplitude decays with increasing angle the fastest in the HP sample, and then in the 3P2 sample, compared to that in the 3P1 and LP samples, indicating a relatively higher interfacial roughness in HP and 3P2 samples. A superlattice peak is observed in the LP sample at $2\theta=8.9º$, corresponding to the bilayer thickness of $d=1.0$ nm (Fig. 1a inset). In the vicinity of this angular position a similar, but broader, peak is observed in both the 3P1 and 3P2 samples, which is absent in the HP sample. These XRR results indicate increasing interfacial roughness from samples LP and 3P1 to 3P2 and HP.

High angle XRD scans (Fig. 1b) reveal (111) texture of the films. The Pd (111) peak at $2\theta=40.1°$ arises from the 20nm Pd seed layer, while the higher angle peak at 41.1° - 41.5° is from the [Co(0.4nm)/Pd(0.6nm)]$_{60}$ multilayer. In the sequence of LP-3P1-3P2-HP, which represents increasing disorder, the Co/Pd (111) peak shifts to lower angle with decreasing intensity while the Pd (111) peak maintains its position and intensity. With increasing structural disorder, the diffraction peak intensity is expected to decrease; for increasing interfacial roughness and interdiffusion of Co and Pd atoms, a shift of the Co/Pd (111) peak towards the alloyed Co$_{0.4}$Pd$_{0.6}$ (111) ($2\theta=40.9°$) is also expected. Note that for the 3P1 and 3P2 samples, the only difference is the order of layer deposition; thus the larger shift of the Co/Pd (111) in 3P2 towards alloyed Co$_{0.4}$Pd$_{0.6}$ (111) indicates larger disorder in the sample, suggesting vertical disorder propagation from the bottom layers (sputtered at 2.7 Pa and 1.6 Pa) to the top (sputtered at 0.7 Pa).

Cross-sectional TEM shows columnar growth in both 3P1 and 3P2 samples (bottom of Fig. 2). Top panels of Figs. 2a and 2b show selected area electron diffraction patterns performed on an area of ~100 nm × 100 nm in the cross-section geometry for sample 3P1 and 3P2, respectively. Sample 3P1 exhibits (111) texture, indicated by arcs in the {111} ring, centered along the Si [001] direction (Fig. 2a). In contrast, in the 3P2 sample the intensity around the {111} ring is more evenly distributed (Fig. 2b), indicating a more random orientation along the beam direction.[37] This is consistent with the effects of larger interfacial roughness and more disorders, which lead to a larger spread of magnetic easy axis, as observed previously in similar Co/Pt films grown at high pressures.[30]

For PNR measurements, the incident neutron beam was polarized to be alternately spin-up or spin-down with respect to a field applied parallel to the sample surface (along the hard axis). The non-spin-flip specular reflectivities, sensitive to the nuclear composition and the in-plane component of the magnetization along the field direction, were measured as functions of scattering vector $Q_z$. Spin-flip scattering (looked for at selected conditions) was found to be negligible, and not considered in the analysis.[18] These measurements were conducted at room temperature as a function of decreasing in-plane applied magnetic field, starting with $H$=3T where the sample reached saturation according to VSM as well as other PNR measurements.[27] The measured 3T PNR reflectivities (discrete data points) for 3P1 and 3P2 with opposite deposition orders are clearly different (Figs. 3a and 3b, respectively), indicating a significant change in their nuclear/magnetic depth profiles. Quantitative information about these profiles was obtained by model-fitting the PNR data, using the Refl1D software package.[38] The experimental data are well-fitted (lines in Figs. 3a and 3b) by the nuclear and magnetic models shown in Figs. 3c and 3d, respectively. While the nuclear depth profiles for the 3P1 and 3P2 samples are almost identical (Fig. 3c), the corresponding magnetization profiles are highly asymmetric (Fig. 3d): in 3P1, the in-plane magnetization at 3T varies from 850 kA/m at the bottom 0.7 Pa region to 365 kA/m at the top 2.7 Pa region. In 3P2, the in-plane magnetization varies from 460 kA/m at the bottom 2.7 Pa layer to 590 kA/m at the top 0.7 Pa region, a much smaller difference. Furthermore, the in-plane magnetization of the 0.7 Pa region in 3P2 (590 kA/m) is much smaller than that in 3P1 (850 kA/m); while that of the 2.7 Pa region in 3P2 (460 kA/m) is only slightly larger than the counterpart in 3P1 (365 kA/m). This contrast indicates that increased deposition pressure strongly suppresses aligned in-plane magnetization, due to a combination of difference in magnetic anisotropy, smaller saturation magnetization, and larger spread of magnetic easy axes.[18, 30] The large difference in the in-plane magnetization of the 0.7Pa region is indicative of the vertical disorder propagation: in 3P1 the 0.7 Pa Co/Pd region is deposited on a pristine Pd seed that was sputtered at 0.7 Pa, while in 3P2 the 0.7 Pa Co/Pd region is deposited on top of two Co/Pd stacks sputtered at 2.7 Pa and 1.6 Pa, respectively, that have rougher interfaces and larger disorders. While less pronounced, the magnetization difference of the 2.7 Pa region is also consistent with disorder propagation: for 3P2, the 2.7 Pa region is deposited on the pristine Pd seed sputtered at 0.7 Pa, and shows higher magnetization than that in the 3P1 counterpart, deposited on Co/Pd sputtered at 1.6 Pa. Additional magnetic-field dependent PNR studies have confirmed that an effective anisotropy gradient is established in the 3P1 sample,[27] while that in 3P2 is much suppressed.

The families of FORCs measured with applied field perpendicular to the film plane are shown in Figs. 4a-4d, whose outer boundaries delineate the major loops. The LP (Fig. 4a) and 3P1 samples (Fig. 4b) both have major loops that are pinched near zero field, characteristic of perpendicular anisotropy systems.[31] The 3P1 sample exhibits slightly larger coercivity and loop squareness than the LP sample, indicating a more stable magnetic configuration. Major loops for the 3P2 (Fig. 4c) and HP (Fig. 4d) both lack the pinched character, but exhibit higher coercivities than the LP and 3P1 (the lower coercivity in 3P2 compared to HP is evidence of the graded magnetic anisotropy). To more clearly analyze the magnetization reversal behavior we turn to the FORC distributions.

The FORC distribution for the LP sample exhibits two distinct features. The first horizontal ridge, at the top of Fig. 4e, corresponds to the initial precipitous drop in magnetization and is caused by the domain nucleation and rapid propagation within the multilayer. For -0.1 T > $\mu_0 H_r$ > -0.4 T there is a shallow plateau in the FORC distribution, corresponding to mostly reversible labyrinth domain expansion and contraction, as has been reported in Co/Pt (Ref. [31]) and Co/Pd (Ref. 20) previously. For $\mu_0 H_r$ < -0.4 T there exists a large negative / positive pair of peaks in the FORC distribution. This feature corresponds to the annihilation of domains and the relatively small spread of this feature in $\mu_0 H_r$ indicates a low density of domain wall pinning sites.

The 3P1 FORC distribution (Fig. 4f) is qualitatively similar to that for the LP sample,

indicating a similar reversal process but with some key differences. Firstly, the horizontal ridge feature is lower in $\mu_0 H_r$, that is, the nucleation of reversal domains occurs at a more negative field. This corresponds to increased stability of the remnant state, which can be attributed to the influence of the magnetically harder regions sputtered at higher pressures. The next distinguishing feature is the plateau in the region -0.2 T $\geq \mu_0 H_r \geq$ -0.4 T. This feature, which is relatively shallow in the LP sample, is more significant in 3P1, indicating the presence of more irreversible switching events during the lateral expansion/contraction of labyrinth domains. This is due to the increased number of pinning sites in the harder, high pressure layers impeding the domain wall motion. Lastly the negative/positive pair of FORC peaks is extended to $\mu_0 H_r <$ -1.0 T, much farther than in the LP sample, even though the major loop seems to be fully saturated in this field range. This is consistent with previous studies of exchange springs[35] and confirms an increased number of pinning sites within the high pressure region which require larger fields to fully saturate.

The HP distribution (Fig. 4h) shows very different behavior from the samples considered so far. Previous studies have shown that the more disordered multilayers have small domain sizes (i.e., more nucleation sites) and the reversal is dominated by domain wall pinning and magnetization rotation.[19] This manifests in the FORC distribution as a broad, single peak with an extended tail feature.

The FORC distribution for sample 3P2 (Fig. 4g) contains elements of both the HP and 3P1 samples. The similarity with the HP sample is clear, confirming that disorder from the high pressure bottom layer is indeed propagating upward into the low pressure region. Close inspection reveals that the 3P2 FORC distribution contains two peaks, similar to 3P1. There is a peak along $\mu_0 H_r$ = -0.2 T which is associated with the nucleation of domains across the multilayer. The broad character of this peak is due to the many disorder sites nucleating domains at different applied fields and impeding the propagation of domain walls. There is another broad peak in the distribution at $\mu_0H_r$ = -0.7 T which corresponds to the rotation/annihilation of domains at the various pinning sites. The saddle-like area between these peaks indicates some domain wall motion between defects and pinning sites. This is likely due to the domain expansion within the softer layers in 3P2.

In summary, we have demonstrated that deposition order of pressure graded perpendicular magnetic anisotropy thin films strongly affects the magnetization reversal mechanisms and magnetic anisotropy gradient. Structural disorder is shown to propagate vertically within the films from the bottom high-pressure region to the top low-pressure region. Reversal in relatively more pristine films (LP, 3P1) is dominated by domain nucleation, propagation, and annihilation. For more disordered films (HP, 3P2) magnetization reversal is dominated by domain wall pinning and magnetization rotation. Graded magnetic anisotropy is effectively established, along with a pronounced depth-dependent magnetization variation, when the relatively more pristine layers were grown at lower sputtering pressure first, compared to samples with the reversed deposition order.

This work has been supported by the NSF (DMR-1008791, ECCS-0925626). Los Alamos National Laboratory is operated by Los Alamos National Security LLC under DOE Contract DE-AC52-06NA25396.

**Figure Captions**

Fig. 1: (a) X-ray reflectivity measurements show the decrease in oscillation amplitude with $2\theta$ for increasing deposition pressure. Inset shows that a superlattice peak is observed for the LP sample but completely absent in the HP sample. 3P1 and 3P2 show degraded superlattice peaks indicating a more disordered bilayer structure. (b) High angle x-ray diffraction scans show a constant Pd seed peak and a shifting of the Co/Pd (111) towards the $Co_{0.4}Pd_{0.6}$ alloy peak position with increasing deposition pressure.

Fig. 2: Selected area electron diffraction patterns of (a) the 3P1 and (b) 3P2 samples (upper panels), along with the corresponding cross-sectional transmission electron microscopy images (lower panels).

Fig. 3: Polarized neutron reflectivities (symbols) and fits (curves) for the spin-up and spin-down channels for the (a) 3P1 and (b) 3P2 samples at H=3T are shown along with (c) the nuclear density profiles used in the fits. (d) In-plane magnetization at H=3 T obtained by fitting the spin dependent reflectivities show contrasting depth profiles.

Fig. 4: Families of FORCs (a-d) and their corresponding distributions (e-h) for [Co(0.4nm) / Pd(0.6nm)]$_{60}$ grown with different pressure profiles.

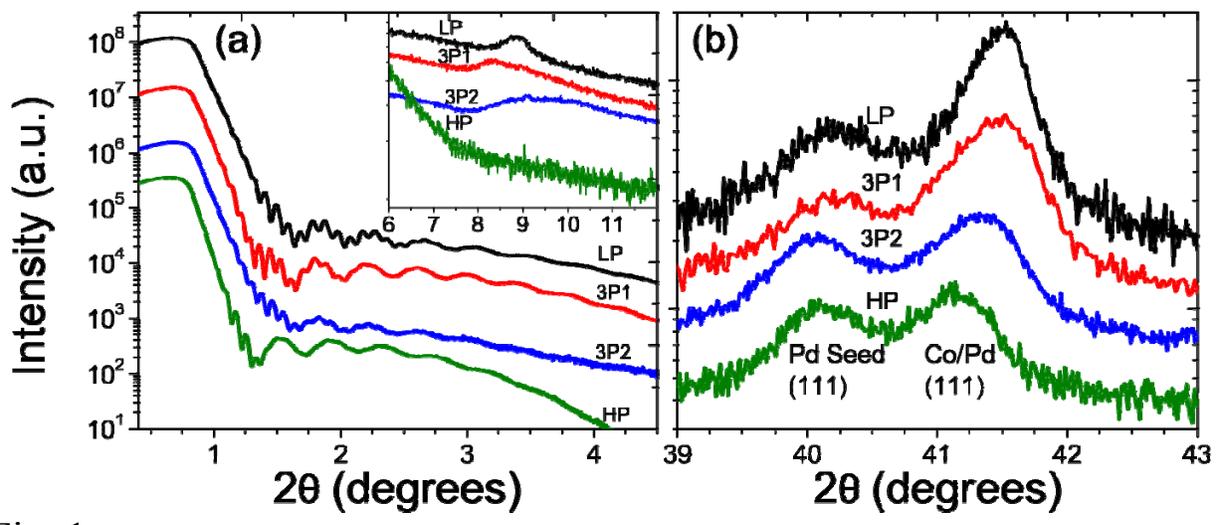

Fig. 1.

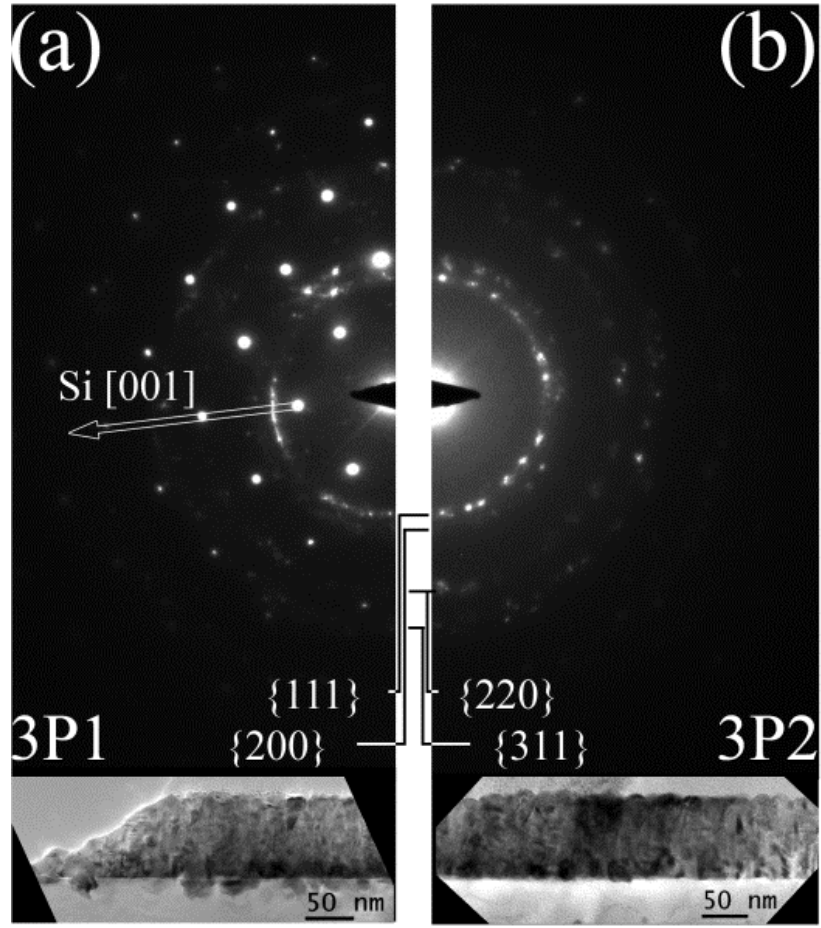

Fig. 2.

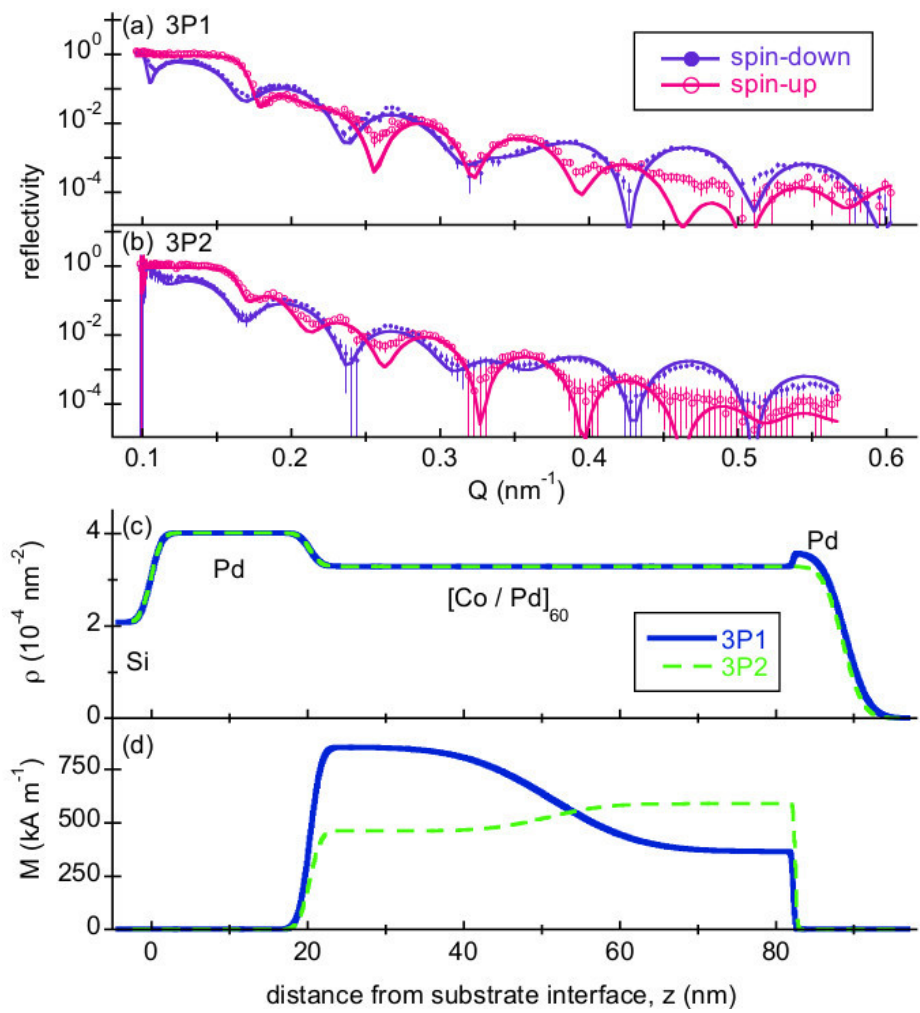

Fig. 3.

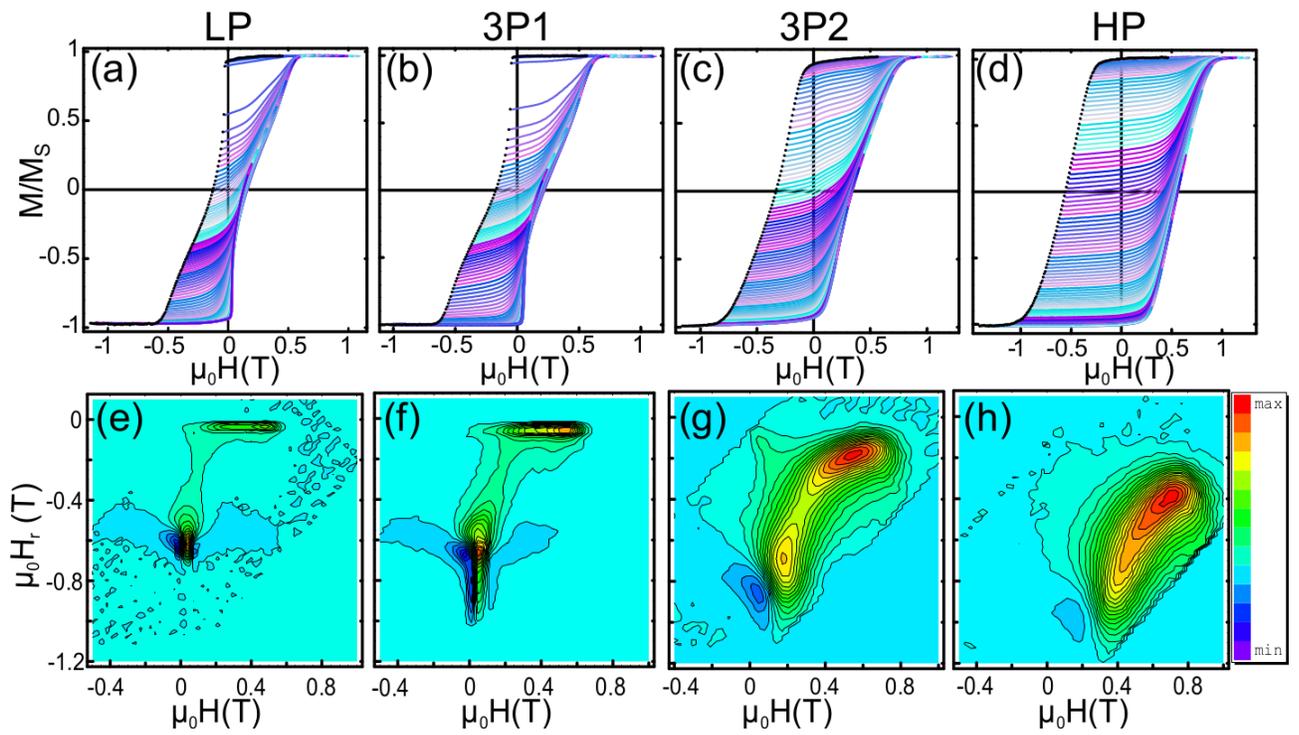

Fig. 4.